\documentclass[twoside]{LCWS11}
\usepackage[latin1]{inputenc}
\usepackage[dvips]{graphicx,epsfig,color}
\usepackage{wrapfig,rotating}
\usepackage{amssymb,amsmath,array}
\usepackage{cite}
\begin{document}
\title{
Testing the Higgs model with triplet fields at the ILC \thanks{This proceedings is based on Ref.~\cite{Kanemura:2011kc}.}} 
\author{Kei Yagyu
\vspace{.3cm}\\
Department of Physics, University of Toyama, 3190 Gofuku, Toyama 930-8555, Japan
}

\maketitle

\begin{abstract}
Higgs triplet fields are introduced various new physics models 
such as the type-II seesaw model, the left-right symmetric model and so on. 
The vertex of a charged Higgs boson and weak gauge bosons, $H^\pm W^\mp Z$, 
appears at the tree level in these models. 
The magnitude of this vertex is proportional to the vacuum expectation value (VEV) of 
Higgs triplet fields. 
We discuss the possibility of measuring this vertex at the ILC and study 
how precisely determine the VEV of the triplet field. 
\end{abstract}

\section{Introduction}
The Higgs sector is the last unknown part of the Standard Model (SM). 
The Higgs sector may not necessarily be the minimal form in the SM. 
Extended Higgs sectors have often been considered in 
various new physics contexts beyond the SM. Therefore, 
determination of the Higgs sector is important to 
obtain a clue to new physics at the TeV scale. 

An important observable to constrain the structure 
of extended Higgs models is the electroweak rho parameter 
$\rho$, whose experimental value is very close to unity.  
This fact suggests that a global 
$SU(2)$ symmetry (custodial symmetry) plays an important 
role in the Higgs sector. 
In the Higgs model which contains complex scalar fields with the isospin $T_i$ and
the hypercharge $Y_i$ as well as real ($Y=0$) scalar fields 
with the isospin $T_i'$, the rho parameter is given at the tree level by
\begin{align}
\rho_{\textrm{tree}}&=\frac{\sum_i\left[|v_i|^2(T_i(T_i+1)-Y_i^2)+u_i^2T_i'(T_i'+1)\right]}{2\sum_i|v_i|^2Y_i^2},\label{rho1}
\end{align}
where $v_i$ ($u_i$) represents the vacuum expectation value (VEV) of the complex (real) scalar field \cite{rho_formula}. 
In the model 
with only scalar doublet fields (and singlets), we obtain 
$\rho_{\text{tree}}=1$ so that the natural extension of the Higgs sector is 
attained by adding extra doublet fields and singlet fields. 
On the other hand, addition of 
the Higgs field with the isospin larger than one half can 
shift the rho parameter from unity at the tree level, whose 
deviation is proportional to the VEVs of these exotic scalar 
fields. The rho parameter, therefore, has been used to 
exclude or to constrain a class of Higgs models. 

A common feature in the extended Higgs models 
is the appearance of physical charged Higgs boson $H^\pm$. 
Hence, we may be able to discriminate each Higgs model through 
the physics of charged Higgs bosons. 
In particular, the $H^\pm W^\mp Z$ vertex can be a useful probe 
of the extended Higgs sector~\cite{Grifols-Mendez,HWZ,HWZ-Kanemura,logan}.  
Assuming that there are several physical charged scalar states 
$H_{\alpha}^\pm$ ($\alpha \geq 2$) and the Nambu-Goldstone modes 
$H_1^\pm$,  the vertex parameter $\xi_\alpha$ in 
$\mathcal{L}=
igm_W \xi_\alpha H_\alpha^+ W^- Z+\textrm{h.c.}$ 
is calculated at the tree level as~\cite{Grifols-Mendez}
\begin{align}
\sum_{\alpha \geq 2} |\xi_\alpha|^2 
&=\frac{1}{\cos^2\theta_W}\left[\frac{2g^2}{m_W^2}
\Big\{\sum_i[T_i(T_i+1)-Y_i^2]|v_i|^2Y_i^2\Big\}-\frac{1}
{\rho_{\textrm{tree}}^2}\right], 
\end{align}
where $\rho_{\rm tree}$ is given in Eq.~(\ref{rho1}). 
A non-zero value of $\xi_\alpha$ appears at the tree level 
only when $H_\alpha^\pm$ comes from an exotic representation such as 
triplets. 
Similarly to the case of the rho parameter, 
the vertex is related to the custodial symmetry. 
In general, this can be independent of the rho parameter. 
Therefore, the measurement of the $H^\pm W^\mp Z$ vertex 
can be a complementary tool to the rho parameter in testing the 
{\it exoticness} of the Higgs sector. 
At LHC, the possibility of measuring the $H^\pm W^\mp Z$ vertex has been 
studied~\cite{WZfusion}.

In this talk, we discuss the possibility of measuring the $H^\pm W^\mp Z$ 
vertex via the process $e^+e^-\to H^\pm W^\mp$ at the International 
Linear Collider (ILC) \cite{Cheung:1994rp,eeHW1,eeHW2} by using the recoil method. 
The feasibility of the signal $e^+e^- \to H^\pm W^\mp\to \ell\nu jj$ is 
analyzed assuming the polarized electron and positron beams and the expected detector performance 
for the resolution of the two-jet system at the ILC, where decay of $H^\pm $ are assumed to be lepton specific~\cite{thdm_Yukawa2}. 
The background events can be reduced to a considerable extent by imposing the kinematic cuts even 
if we take into account the initial state radiation (ISR).

\section{The $H^\pm W^\mp Z$ vertex and the process $e^+e^-\to H^\pm W^\mp$}
%

\begin{wrapfigure}{r}{0.45\columnwidth}
\centerline{\includegraphics[width=0.4\columnwidth]{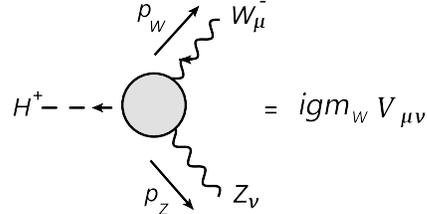}}
\caption{The $H^\pm W^\mp Z$ vertex.}\label{HWZ_proceedings}
\end{wrapfigure}

The $H^\pm W^\mp Z$ vertex
\footnote{The $H^\pm W^\mp \gamma$ vertex vanishes at the tree level due to the $U(1)_{\rm em}$ gauge invariance 
in any extended Higgs models~\cite{HWZ}. 
} is defined in FIG. \ref{HWZ_proceedings}, where   
$V^{\mu\nu}$ is expressed in terms of the form factors 
$F_{HWZ}$, $G_{HWZ}$ and $H_{HWZ}$ as 
\begin{align}
V^{\mu\nu}=F_{HWZ}g^{\mu\nu}+G_{HWZ}\frac{p_W^\mu p_Z^\nu}{m_W^2}
       +iH_{HWZ}\frac{p_{W\rho} p_{Z\sigma}}{m_W^2}\epsilon^{\mu\nu\rho\sigma},
\label{HWV}
\end{align}
with $\epsilon_{\mu\nu\rho\sigma}$ being the anti-symmetric tensor, and $p_Z^\mu$ and $p_W^\mu$ being the 
outgoing momenta of $Z$ and $W$ bosons, respectively. 
The form factors $G_{HWZ}$ and $H_{HWZ}$ 
are related to the coefficients of the dimension five operator in the Lagrangian~\cite{HWZ,HWZ-Kanemura}, 
while $F_{HWZ}$ is related to those of the dimension three operator, so that only $F_{HWZ}$ may appear at the tree level. 
Therefore, the dominant contribution to the $H^\pm W^\mp Z$ vertex 
is expected to be from $F_{HWZ}$.

In the Higgs model with only doublet scalar fields (plus singlets) 
all the form factors including $F_{HWZ}$ vanish 
at the tree level~\cite{Grifols-Mendez}, because of the 
custodial invariance in the kinetic term. 
The form factors  $F_{HWZ}$, $G_{HWZ}$ and $H_{HWZ}$ 
are generally induced at the loop level. 
On the other hand, in models with Higgs triplet fields, 
the $H^\pm W^\mp Z$ vertex appears at the tree level.  
In the model with an isospin doublet field ($Y=1/2$) 
and either an real triplet field $\eta$ ($Y=0$) 
or an additional complex triplet field $\Delta$ ($Y=1$),  
concrete expressions for the tree-level formulae for  
$|F_{HWZ}|^2$ and that of $\rho_{\textrm{tree}}$ are 
shown in TABLE~\ref{f_models}, 
where $v$, $v_\eta$ and $v_\Delta$  are respectively 
VEVs of the doublet scalar field and 
the additional triplet scalar fields $\eta$ and $\Delta$. 
\begin{table}[t] 
\begin{center}
{\renewcommand\arraystretch{1.2}
\begin{tabular}{|c||c|c|c|}\hline
Model & SM with $\eta$  ($Y=0$)& SM with $\Delta$ ($Y=1$) 
& the GM model \\\hline\hline
$|F_{HWZ}|^2=$&$\frac{4v^2 v^{2}_\eta}{\cos^2\theta_W(v^2+4v^{2}_\eta)^2}$
&$\frac{2v^2 v^{2}_\Delta}{\cos^2\theta_W(v^2+2v^{2}_\Delta)^2 }$& 
$\frac{4 v_\Delta^2}{\cos^2\theta_W(v^2+4v_\Delta^2)}$ 
\\\hline 
$\rho_{\text{tree}}=$&$1+\frac{4v^{2}_\eta}{v^2}$
&$\frac{1+2\frac{v^{2}_\Delta}{v^2}}{1+4\frac{v^{2}_\Delta}{v^2}}$&$1$ \\\hline
\end{tabular}}
\caption{The tree-level expression 
for $F_{HWZ}$ and rho parameter in the model with a real triplet field $\eta$, that with a complex triplet field $\Delta$
and the Georgi-Machacek (GM) model~\cite{Georgi:1985nv}.}
\label{f_models}
\end{center}
\end{table}
These triplet scalar fields also contribute to 
the rho parameter at the tree level, so that their VEVs 
are constrained by the current rho parameter data,  
$\rho_{\exp}=1.0008^{+0.0017}_{-0.0007}$ \cite{rho_exp}; i.e., 
$v_\eta \lesssim 6$ GeV for the VEV of $\eta$, 
and $v_\Delta \lesssim 8$ GeV for that of $\Delta$ 
(95 \% CL). 
We note that in order to obtain the similar accuracy to the rho 
parameter data by measuring the $H^\pm W^\mp Z$ vertex, 
the vertex  has to be measured with the detectability to    
$|F_{HWZ}|^2 \sim {\cal O}(10^{-3})$.  

Finally, we mention the model with $\eta$ and $\Delta$ in addition to the SM, 
which is proposed by Georgi-Machacek and 
Chanowiz-Golden~\cite{Georgi:1985nv, Gunion:1989ci,AK_GM}. 
In this model, an alignment of the VEVs for $\eta$ and $\Delta$ 
are introduced ($v_\eta= v_\Delta/\sqrt{2}$), 
by which the Higgs potential is invariant under   
the custodial symmetry at the tree level. 
Physical scalar states in this model can be classified 
using the transformation property against 
the custodial symmetry; i.e., 
the five-plet, the three-plet and the singlet. 
Only the charged Higgs boson from the five-plet state has 
the non-zero value of $F_{HWZ}$ at the tree level. 
Its value is proportional to $v_\Delta$. 
However, the value of $v_\Delta$ is not strongly constrained 
by the rho parameter data, because 
the tree level contribution to the rho parameter 
is zero due to the custodial symmetry: 
see TABLE~\ref{f_models}. 
Consequently, the magnitude of $|F_{HWZ}|^2$ 
can be of order one.\\\\
\noindent
%

\begin{figure}[t]
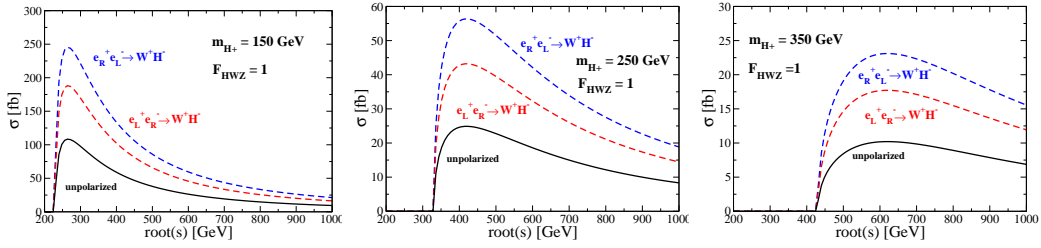

\begin{center}
\includegraphics[width=45mm]{cs1_proc.eps}
\includegraphics[width=45mm]{cs2_proc.eps}
\includegraphics[width=45mm]{cs3_proc.eps}
\caption{The total cross section of $e^+e^-\to W^+H^-$ as a function of $\sqrt{s}$ 
in the case of $F_{HWZ}=1$.}
\label{mch150_fz1_02}
\end{center}
\end{figure}

The process $e^+ e^-\to Z^* \to H^- W^+$  is directly related to the $H^\pm W^\mp Z$ vertex. 
The helicity specified cross sections of the process are calculated as a function of 
the center-of-mass energy $\sqrt{s}$ and the helicity of the electron $\tau$ \cite{eeHW1}; 
\begin{align}
\sigma(s; \tau)=
\frac{1}{32\pi s}\beta\left(\frac{m_{H^\pm}^2}{s},\frac{m_W^2}{s}\right)
\int_{-1}^{1}d\cos\theta|\mathcal{M}(\tau)|^2,
\end{align}
where $\theta$ and $m_{H^+}$ are the angle between the momentum of $H^\pm$ and the 
beam axis and the mass of $H^{\pm}$. The function $\beta(x,y)$ is 
\begin{align}
\beta(x,y)=\sqrt{1+x^2+y^2-2xy-2x-2y}. 
\end{align}
The squired amplitude $|\mathcal{M}(\tau)|^2$ can be written as 
\begin{align}
|\mathcal{M}(\tau)|^2
=g^2C_Z^2\frac{|F_{HWZ}|^2}{(s-m_Z^2)^2}\left[\frac{\sin^2\theta}{4}
(s+m_W^2-m_{H^\pm}^2)^2+sm_W^2(\cos^2\theta+1)\right],\label{amp2}
\end{align}
where the form factors $G_{HWZ}$ and $H_{HWZ}$ are taken to be zero and
\begin{align}
C_Z=
\frac{g}{\cos\theta_W}(T_e^3+\sin^2\theta_W),
&\quad T_e^3=
\left\{
\begin{array}{ll}
-1/2 &\text{for }\tau=-1, \\
0    &\text{for }\tau=+1 
\end{array}\right. .
\end{align}
In FIG.~\ref{mch150_fz1_02}, we show that the 
$\sqrt{s}$ dependence of the helicity dependent and the helicity averaged cross sections. 

\section{The signal and background analysis}

We investigate the possibility of measuring 
the $H^\pm W^\mp Z$ vertex by using a recoil method~\cite{recoil} at the ILC. 
In order to identify the process, 
we consider the hadronic decays $W\to jj$ 
instead of the leptonic decay of the produced $W$ boson.  
The recoiled mass of $H^\pm$ is given in terms of the 
two-jet energy $E_{jj}$ and 
the two-jet invariant mass $M_{jj}$ as 
\begin{align}
m_{\text{recoil}}^2(jj)=s-2\sqrt{s}E_{jj}+M_{jj}^2.\label{recoil2}
\end{align}
It is clear that the detector performance for 
the resolution of two jets is crucial in such an analysis. 
In particular, the jets from the $W$ boson in the signal process 
have to be precisely measured in order to be separated with those from the 
$Z$ boson in the background process. 
At the ILC, the resolution for the two jet system with the energy $E$ 
in the unit of GeV is expected to be $\sigma_E = 0.3 \sqrt{E}$ GeV, 
by which the background 
from $Z\to jj$ can be considerably reduced.  
We here adopt the similar value for $\sigma_E$=3 GeV 
in our later analysis. 

At the ILC, the polarized electron and positron beams can be used, by 
which the background from the $W$ boson pair production process can be 
reduced. We here use the following beams polarized as  
\begin{align}
\frac{N_{e_R^-}-N_{e_L^-}}{N_{e_L^-}+N_{e_R^-}} =0.8,\quad 
\frac{N_{e_L^+}-N_{e_R^+}}{N_{e_L^+}+N_{e_R^+}} =0.5,\label{porl}
\end{align}
which are expected to be attained at the ILC~\cite{ILCTDR2007}, 
where 
$N_{e_{R,L}^-}$ and $N_{e_{R,L}^+}$ are numbers of right- (left-) handed electron and 
positron in the beam flux per unit time. 

First, we discuss the case without the effect of the ISR, and after that 
we discuss the case with the ISR. 
The size of the signal cross section is determined by 
$\sqrt{s}$, $m_{H^\pm}^{}$ and  $F_{HWZ}$. 
In order to examine the possibility of constraining $|F_{HWZ}|^2$, 
we here assume that $m_{H^\pm}$ is already 
known with some accuracy by measuring the other processes at the LHC or at the ILC. 
Then $|F_{HWZ}|^2$ is the only free parameter in the production cross section.  

In order to perform the signal and background analysis, 
we assume that the decay of the produced $H^\pm$ is 
lepton specific; i.e., $H^\pm \to \ell \nu$ where $\ell$ is either 
$e$, $\mu$ or $\tau$.  
The final state of the signal is then $e^+e^- \to H^\pm W^\mp \to \ell \nu jj$. 
The main backgrounds come from the $W$ boson pair production process 
$e^+e^- \to W^+W^-$ and the single $W$ production processes 
$e^+e^- \to Z^*/\gamma^*\to W^\pm jj$ and $e^+e^- \to Z^*/\gamma^*\to W^\pm \ell^\mp\nu$. 
For the $e^\pm \nu jj$ final state, additional processes 
$e^+e^-\to e^\pm \nu W^{\mp*}$, $e^+e^-\to e^\pm \nu W^{\mp*}Z^*$ and $e^+e^-\to e^\pm \nu W^{\mp*}\gamma^*$ 
can also be significant backgrounds. 
In addition, we take into account the processes with the final 
state of $\ell\ell jj$. They can be backgrounds if one of the outgoing leptons escapes from 
the detection at the detector. We assume that 
the efficiency for lepton identification is 90~\%.

We impose the basic cuts for all events such as 
\begin{align}
10^\circ < A_j < 170^\circ,\quad 5^\circ < A_{jj} < 175^\circ,
\quad 10\; \textrm{GeV}< E_{jj}, \label{basic}
\end{align}
where $A_j$ is the angle between a jet and the beam axis, 
$A_{jj}$ is the angle between the two jets 
and $E_{jj}$ is the energy of the two jets. 
In the numerical evaluation, we use CalcHEP~\cite{CalcHEP}.
After the basic cuts, the cross section for the signal is 
0.15 fb and that for the background is 1.2 pb, 
where we set $\sqrt{s}=300$ GeV, $m_{H^\pm}=150$ GeV and $|F_{HWZ}|^2=10^{-3}$. 
\begin{table}[t]
\begin{center}
{\renewcommand\arraystretch{1.2}
\begin{tabular}{|l||c|c|c|c||c|c|}\hline
 &Basic & $M_{jj}$ \hspace{5mm}  &  $p_T^{jj}$\hspace{5mm}  & $E_{jj}$ \hspace{5mm} & $\cos\theta_{\text{lep}}$ & $M_{\ell \nu}$\hspace{5mm}  \\\hline\hline
Signal~(fb) & 0.15 &0.14&8.9$\times 10^{-2}$  &8.9$\times 10^{-2}$ &7.0$\times 10^{-2}$&7.0$\times 10^{-2}$\\\hline
$\ell\nu jj$ background~(fb)&820 &720 &120 &7.4 &1.5 &8.0$\times 10^{-1}$\\\hline
$\ell\ell jj$ background~(fb) &330 &5.2 &3.0$\times 10^{-1}$ &2.5$\times 10^{-2}$  &1.2$\times 10^{-2}$ &6.4$\times 10^{-3}$\\\hline
$S/\sqrt{B}$  &0.14&0.16&0.26&1.0&1.8&2.5\\\hline
\end{tabular}}
\caption{The results without the ISR. 
The signal and the backgrounds cross sections are shown for $\sqrt{s}=300$ GeV.  
For the signal, $m_{H^\pm}$ is 150 GeV and  
$|F_{HWZ}|^2$ is taken to be 10$^{-3}$. 
For the $\ell\ell jj$ processes, the misidentity rate of one 
of the leptons is assumed to be 0.1. 
The signal significance $S/\sqrt{B}$ are evaluated for 
the integrated luminosity to be 1~ab$^{-1}$.}
\label{result1}
\end{center}
\end{table}
\begin{table}[t]
\begin{center}
{\renewcommand\arraystretch{1.2}
\begin{tabular}{|l||c|c|c|c||c|c|}\hline
 &Basic & $M_{jj}$ \hspace{5mm}  &  $p_T^{jj}$\hspace{5mm}  & $E_{jj}$ \hspace{5mm} & $\cos\theta_{\text{lep}}$ & $M_{\ell \nu}$\hspace{5mm}  \\\hline\hline
Signal~(fb) &0.14 &0.13 &6.9$\times 10^{-2}$ &6.6$\times 10^{-2}$ &5.2$\times 10^{-2}$ &5.1$\times 10^{-2}$ \\\hline
$\ell\nu jj$ background~(fb)  &810 &720 &130 &13 &6.2 &6.7$\times 10^{-1}$ \\\hline
$\ell\ell jj$ background~(fb) &360 &4.6 &0.29 &3.4$\times 10^{-2}$ &2.1$\times 10^{-2}$ &5.5$\times 10^{-3}$\\\hline
$S/\sqrt{B}$ &0.13 &0.15&0.19&5.8$\times 10^{-1}$&6.6$\times 10^{-1}$&2.0\\\hline
\end{tabular}}
\caption{The results with the ISR. 
The signal and the backgrounds cross sections are shown for $\sqrt{s}=300$ GeV.  
For the signal, $m_{H^\pm}$ is 150 GeV and  
$|F_{HWZ}|^2$ is taken to be 10$^{-3}$. 
For the $\ell\ell jj$ processes, the misidentity rate of one 
of the leptons is assumed to be 0.1. 
The signal significance $S/\sqrt{B}$ are evaluated for 
the integrated luminosity to be 1 ab$^{-1}$.}
\label{result2}
\end{center}
\end{table}
In order to improve the signal over background 
ratio, we impose additional kinematic cuts. 
The two jets come from the $W$ boson for the signal, so that 
the following invariant mass cut is useful to reduce the backgrounds; 
\begin{align}
m_W-2\sigma_E < M_{jj} < m_W+2\sigma_E.  \label{mjj}   
\end{align}

\begin{figure}[t]
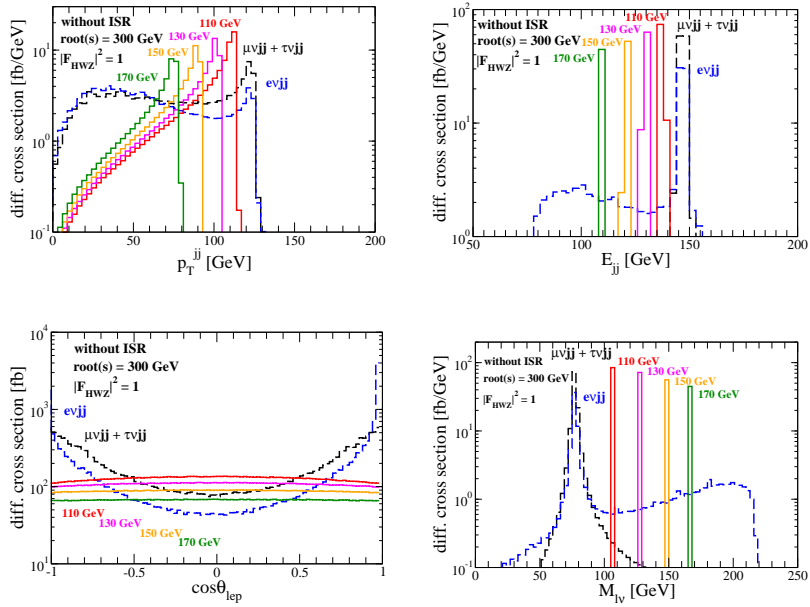

\begin{center}
\includegraphics[width=50mm]{dist1_proc.eps}\hspace{5mm}
\includegraphics[width=50mm]{dist2_proc.eps}\\
\vspace{6mm}
\includegraphics[width=50mm]{dist3_proc.eps}\hspace{5mm}
\includegraphics[width=50mm]{dist4_proc.eps}
\caption{Distributions of the signal 
for $m_{H^\pm}=110$, 130, 150 and 170 GeV as well as the backgrounds 
after $M_{jj}$ cut in Eq.~(\ref{mjj}) without the ISR  
as a function of the transverse momentum 
$p_T^{jj}$ (upper left), the energy of the $jj$ system (upper right), 
the angle $\theta_{\rm lep}$ of a charged lepton with the beam 
axis (lower left), and the invariant mass $M_{\ell\nu}$ of the charged lepton and the missing momentum 
in the final state (lower right).  
$|F_{HWZ}|^2$ is taken to be~1.}
\label{distr1}
\end{center}
\end{figure}
In FIG.~\ref{distr1}, the differential cross sections of the signal and the 
backgrounds are shown for the events after the $M_{jj}$ cut in Eq.~(\ref{mjj}) 
as a function of the transverse momentum $p_T^{jj}$, 
the energy of the $jj$ system, the angle $\theta_{\rm lep}$ of a charged 
lepton with the beam axis, and 
the invariant mass $M_{\ell\nu}$ of the charged lepton and the missing momentum 
in the final state.  
For the signal, the results are shown 
for $|F_{HWZ}|^2=1$ with $m_{H^+}$ to be 110, 130, 150 and 170 GeV. 
%

In the following, we discuss the case with $m_{H^\pm}=150$ GeV. 
According to FIG.~\ref{distr1}, we impose the following 
four kinematic cuts sequentially: 
\begin{align}
  75\text{ GeV} <p_T^{jj}<100\text{ GeV}\quad\text{and}\quad 115\text{ GeV} <E_{jj}<125\text{ GeV},  
\end{align}
for the $jj$ system in the final state.  
In TABLE~\ref{result1}, the resulting values for the cross sections 
for the signal and backgrounds are shown in each step of the cuts. 
For $|F_{HWZ}|^2=10^{-3}$, the signal significance reaches 
to ${\cal O}(1)$ assuming the integrated luminosity of 
1 ab$^{-1}$.  

Until now, we have imposed the cuts on the $jj$ system, 
and no information from the $\ell \nu$ system has been used.  
Here, in order to further improve the signal significance,   
we impose new cuts related to the $\ell \nu$ system in order, which 
are determined from FIG.~\ref{distr1}; 
\begin{align}
|\cos\theta_{\text{lep}}| < 0.75\quad\text{and}\quad 144\text{ GeV} <M_{\ell\nu}<156\text{ GeV}. \label{cut4} 
\end{align}
As shown in TABLE~\ref{result1}, for $|F_{HWZ}|^2=10^{-3}$  
the signal significance after these cuts can reach 
to $S/\sqrt{B} \simeq 2.5$ 
assuming the integrated luminosity of $1$ ab$^{-1}$.

Next let us see how this results can be changed by including the ISR. 
We use the beam parameters which are defined in CalcHEP~\cite{CalcHEP} as the default value\footnote{
We have confirmed that the results are almost unchanged 
even when we use the values given in Ref.~\cite{ILCTDR2007}.}. 
The biggest change can be seen in the $E_{jj}$ distribution. 
The background events in the $E_{jj}$ distribution originally located at the point just below 150 GeV 
in the case without the ISR,  
which corresponds to the $W$ boson mass. 
This 
tends to move in the lower $E_{jj}$ regions, 
so that the signal over background ratio becomes worse. 
In TABLE~\ref{result2}, the resulting values for the cross sections 
for the signal and backgrounds are shown in each step of the cuts. 
Consequently, the signal significance after all the cuts is smeared 
from $2.5$ to $2.0$. 
We stress that even taking the ISR into account, 
the $H^\pm W^\mp Z$ vertex 
with $|F_{HWZ}|^2 > 10^{-3}$ can be  excluded with 95\%~CL. 

In the above analysis, we have assumed the lepton specific $H^\pm$ scenario, where 
$H^{\pm}$ decay into $\ell\nu$, and we have not specified the branching fractions of 
$B(H^\pm\to e^\pm \nu)$, $B(H^\pm\to \mu^\pm \nu)$ and $B(H^\pm\to \tau^\pm \nu)$ which 
depend on details of each Higgs model. 
If we assume $B(H^\pm\to e^\pm\nu)=1$, 
the signal cross section does not change from the result shown in Table~\ref{result2}, 
while the background becomes 70 \% of all the $\ell\nu jj$ background as 
evaluated from Table~\ref{result2}. 
As the result, the signal significance $S/\sqrt{B}$ becomes about 2.4 in the case with the ISR for $|F_{HWZ}|^2=10^{-3}$. 
Similarly, If we assume $B(H^\pm\to \mu^\pm\nu)=1$, 
the signal cross section does not change from the result shown in Table~\ref{result2}, 
while the background becomes 15 \% of all the $\ell\nu jj$ background. 
Thus, the signal significance $S/\sqrt{B}$ becomes about 5.0 in the case with the ISR for $|F_{HWZ}|^2=10^{-3}$. 

Finaly, we discuss the upper bound for the VEVs of the triplet fields. 
The constraint for $|F_{HWZ}|^2$ can be translated into that for the VEVs of the triplet fields (see Table~\ref{f_models}). 
The upper bound for the VEVs of the triplet field are listed in Table~\ref{constraint} 
by the constraint from the rho parameter data with 95\% CL 
and for $|F_{HWZ}|^2 <10^{-3}$. 
\begin{table}[t] 
\begin{center}
{\renewcommand\arraystretch{1.2}
\begin{tabular}{|c||c|c|c|}\hline
Model & SM with $\eta$  ($Y=0$)& SM with $\Delta$ ($Y=1$) 
& the GM model \\\hline\hline
$\rho_{\text{exp}}$ with 95\% CL&$v_\eta <$6 GeV
&$v_\Delta <$8 GeV& - 
\\\hline 
$|F_{HWZ}|^2 <10^{-3}$&$v_\eta <$3 GeV&$v_\Delta <$4 GeV&$v_\Delta <$3 GeV \\\hline
\end{tabular}}
\caption{The upper bound for the VEVs of the triplet field from the constraint by the rho parameter data with 95\% CL and $|F_{HWZ}|^2 <10^{-3}$. }
\label{constraint}
\end{center}
\end{table}
\section{Conclusion}

We have discussed the possibility of measuring  the $H^\pm W^\mp Z$ vertex 
at the ILC. 
The vertex is important to understand the exoticness of the Higgs sector, 
so that the combined information of this vertex with the rho parameter 
provides a useful criterion to determine the structure of the extended Higgs sector. 
Assuming that the decay of the charged Higgs bosons is lepton specific, 
the feasibility of the vertex is analyzed by using the recoil method via the process 
$e^+e^- \to H^\pm W^\mp \to \ell\nu jj$ 
with the parton level simulation for the background reduction. 
We have found that the vertex with  $|F_{HWZ}|^2 \geq {\cal O}(10^{-3})$ can be 
excluded  with the 95\% confidence level when $120$-$130$ GeV $ < m_{H^\pm} < m_W+m_Z$.  
%
The measurement of the $H^\pm W^\mp Z$ vertex with 
$|F_{HWZ}|^2 \geq {\cal O}(10^{-3})$ gives a precise information 
for the Higgs sector, whose accuracy is similar to that of the rho parameter.

\section*{Acknowledgments}
I would like to thank Shinya Kanemura and Kazuya Yanase for fruitful collaborations. 
This work was supported by JSPS Fellow.


\begin{footnotesize}


\end{footnotesize}


\end{document}